\documentclass[twocolumn,showpacs,prl,floatfix,amssymb]{revtex4}

\usepackage{graphicx}

\begin{document}

\newlength{\plotwidth}          
\setlength{\plotwidth}{8.3cm}   

\title{Flux-quantum-modulated Kondo conductance in a multielectron quantum dot}

\author{C.~F\"uhner}
\email{fuehner@nano.uni-hannover.de}
\author{U.~F.~Keyser}
\author{R.~J.~Haug}
\affiliation{Institut f\"ur Festk\"orperphysik, Universit\"at
Hannover, Appelstr. 2, 30167 Hannover, Germany}

\author{D.~Reuter}
\author{A.~D.~Wieck}%
\affiliation{Lehrstuhl f\"u{}r Angewandte Festk\"o{}rperphysik,
Ruhr-Universit\"at Bochum, 44780 Bochum, Germany}

\date{\today}

\begin{abstract}
We investigate a lateral semiconductor quantum dot with a large
number of electrons in the limit of strong coupling to the leads.
A Kondo effect is observed and can be tuned in a perpendicular
magnetic field.  This Kondo effect does not exhibit Zeeman
splitting. It shows a modulation with the periodicity of one flux
quantum per dot area at low temperatures.  The modulation leads to
a novel, strikingly regular stripe pattern for a wide range in
magnetic field and number of electrons.
\end{abstract}


\maketitle


The Kondo effect \cite{Kondo-64} in semiconductor quantum dots
\cite{Kouwenhoven-97} has been the subject of numerous theoretical
and experimental investigations in recent years. Originally, Kondo
developed his theory to explain the increased resistivity in bulk
metals due to magnetic impurities at low temperatures.  Later,
this theory was also applied to quantum dots
\cite{Glazman-88,Ng-88}, leading to first successful experiments
in semiconductor dots which allowed a controlled and detailed
study of the Kondo phenomenon
\cite{Kondo-Alle}.  These early experiments were interpreted in
terms of the ordinary Anderson impurity model \cite{Anderson-61},
describing the interaction of a singly occupied, spin-degenerate
level --- realized within the quantum dot --- with conduction band
electrons.  Deviations were attributed to nearby levels present in
real dots \cite{Yeyati-99}.  Subsequently, the quantum dots have
been tuned into more complex regimes, showing Kondo behavior
beyond the simple spin-1/2 Anderson impurity model: For a quantum
dot with many electrons the occurrence of Kondo physics depends on
the exact form of the multi-electron spectrum including Hund's
rule \cite{Schmid-00}. In a magnetic field, the ground state of
such a system and thus the ability to show a Kondo effect is
modified. Unpaired spin-configurations at the edge may lead to a
Kondo effect \cite{Keller-01,Tejedor-01}. Furthermore, multiple
correlated electrons on the dot may couple in a $S \geq 1$ state
which may also exhibit Kondo physics. This was demonstrated in
magnetically tuned degeneracies between singlet and triplet states
\cite{Sasaki-00} (also observed in carbon nanotubes
\cite{Nygard-00}) and is now well understood
\cite{Giuliano-00,Pustilnik-Alle,Eto-00}. Another deviation from
the Anderson model can be observed when several energy levels of
the dot come into play. This happens when either the coupling of
the dot to the leads is not small compared to the level spacing
anymore \cite{Chudnovskiy-01,Pustilnik-01,Hofstetter-02,Wiel-02}
or when the dot is tuned close to a Coulomb resonance
\cite{Costi-94,Goldhaber-98b}. The system then enters the mixed
valence regime.

\begin{figure} 
  \begin{center}
  \resizebox{\plotwidth}{!}{\includegraphics{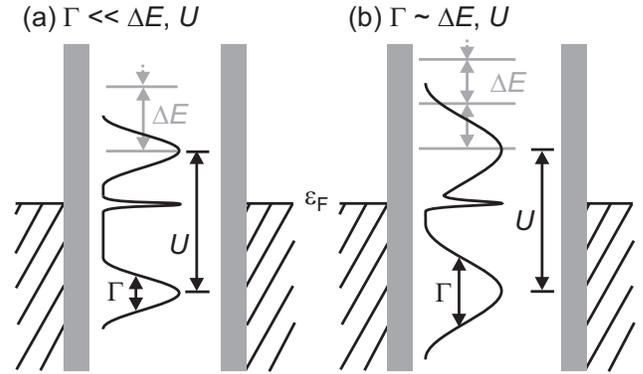}}
  \end{center}
  \caption{%
    Schematic energy diagrams of quantum dots with charging energy
    $U$, level spacing $\Delta E$ and tunnel coupling $\Gamma$.
    (a) A dot in the Kondo regime with the characteristic narrow
    resonance in the density of states at the Fermi energy $\epsilon_F$.
    (b) The situation in our dot, where $U$, $\Delta E$ and $\Gamma$
    are of similar magnitude.
    }
  \label{figRegime}
\end{figure}

In this work, we study Kondo physics in a rather large lateral
quantum dot containing many electrons.  While the energies
involved in tunneling through dots resembling the Anderson
impurity model can be depicted like in Fig. \ref{figRegime} (a),
the situation in our dot is more like in Fig. \ref{figRegime} (b).
Due to the larger size of our dot, the internal level spacing
$\Delta E$ is comparably small, whereas the coupling $\Gamma$ to
the leads is strong so that $\Gamma \approx \Delta E$.  Kondo as
well as mixed valence effects should play a role. In addition, the
large number of electrons in our dot provides electronic ground
states complex enough to allow a variety of Kondo effects like a
singlet-triplet Kondo effect or even more complicated ones.  These
ground states are tuned in a magnetic field.


\begin{figure} 
  \begin{center}
  \resizebox{\plotwidth}{!}{\includegraphics{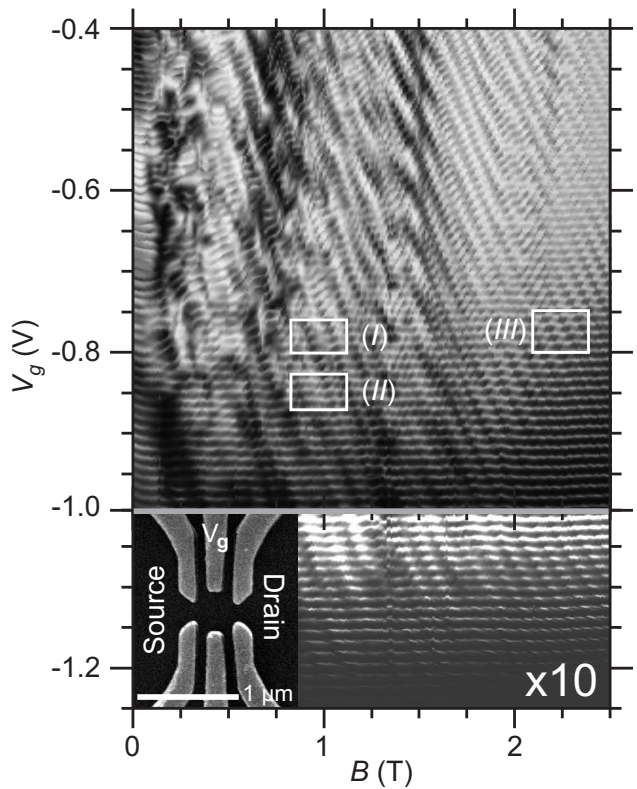}}
  \end{center}
  \caption{%
    Grey scale plot of the linear conductance $G$ as a
    function of plunger gate voltage and perpendicular magnetic field, black
    corresponds to zero conductance and white to
    $G=1.8$~$e^2/h$.  For $V_g < -1$~V, the contrast has been enhanced by
    a factor of 10.  A striking, regular diagonal stripe
    pattern is visible especially for magnetic fields $B>1$~T.
    Inset: SEM picture of the Cr/Au gates used to define the
    quantum dot. The gate marked with $V_g$ is the plunger gate, all other gates
    are kept at constant voltages.
    }
  \label{figStreifen}
\end{figure}

We fabricated our sample from a modulation doped GaAs/AlGaAs
heterostructure which forms a two-dimensional electron gas 57~nm
below the surface with an electron density of $n = 3.7 \cdot
10^{15}$~m$^{-2}$ and a low temperature mobility of $\mu =
130$~m$^2$/Vs. After etching and contacting a standard Hall bar
structure using optical lithography, we applied electron beam
lithography to pattern six metallic gate electrodes (6~nm Cr,
25~nm Au) across the Hall bar (inset in Fig. \ref{figStreifen}).
From the SEM image a geometric dot diameter of 380~nm is deduced.
Taking into account the depletion length of our split gates, we
obtain an electronic diameter of $d_{el} \approx 250$~nm resulting
in a dot with $N \approx 180$ electrons.  From our dot geometry we
estimate a confinement-induced single particle level spacing of
$\Delta E \approx 2 \hbar^{2}/m^*r^2 = 150$~$\mu$eV. Differential
conductance measurements were carried out in a $^3$He-$^4$He
dilution refrigerator with a base temperature of 20~mK, using
standard lock-in technique. From the saturation of peak widths in
temperature dependent Coulomb blockade measurements we determine
the minimal electronic temperature to be $T_0 \lesssim 70$~mK.


The quantum dot is defined by application of negative voltages to
the gate electrodes:  Slightly asymmetric tunnel barriers are
created by applying -1.084~V to the left and -1.220~V to the right
gate pair. The lower gate in the middle of the structure is kept
constant at -1.084~V while the upper one is used as a plunger gate
to control the electrostatic potential, increasing the number of
electrons on the dot by 75 in varying $V_g$ from $-1.250$~V to
$-0.400$~V. Due to its proximity to the tunnel barriers the
plunger gate strongly influences the coupling of the dot to the
leads. For voltages around $V_g \sim -1.2$~V our dot is in the
Coulomb blockade regime with high barrier resistances $R_T \gg
h/e^2$.  For $B=0$~T we find a charging energy of $U \approx
600$~$\mu$eV from Coulomb blockade diamonds and an intrinsic line
width of $\Gamma \approx 100$~$\mu$eV estimated from temperature
dependent measurements of the Coulomb blockade peak width.  Near
$V_g \sim -1.0$~V a distinct Kondo resonance appears in several
consecutive Coulomb blockade diamonds.  In this regime, we
estimate our line width as approximately $\Gamma \approx
250$~$\mu$eV and the charging energy $U \approx 500$~$\mu$eV at
$B=0$~T. From excitation spectroscopy measurements we extract a
level spacing of $\Delta E \approx 100$~$\mu$eV in rough agreement
with the estimation presented above.


\begin{figure} 
  \begin{center}
  \resizebox{\plotwidth}{!}{\includegraphics{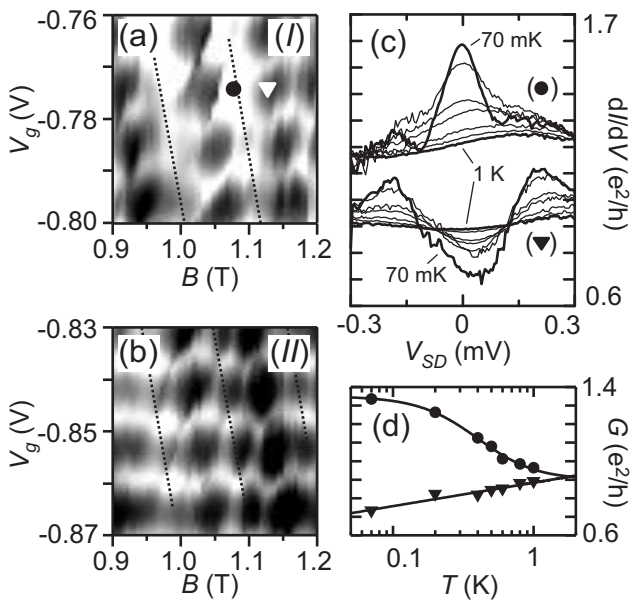}}
  \end{center}
  \caption{%
    (a) and (b) A more detailed view into the regions (I) and (II)
    marked in Fig.~\ref{figStreifen} with the extrapolated stripe
    positions highlighted by dotted lines
    (black corresponds to $G=0.4$~$e^2/h$ and white to
    $1.2$~$e^2/h$).
    (c) Differential conductance versus $V_{SD}$ in high ($\bullet$) and low
        ($\blacktriangledown$) conductance regions marked in (a) for
        temperatures $T=70, 200, 400, 500, 600, 800$ and $1000$~mK.
    (d) Temperature dependence of linear conductance ($V_{SD}=0$~V) at
        the positions marked in (a) and corresponding fits.
    }
  \label{figWaben}
\end{figure}

\begin{figure} 
  \begin{center}
    \resizebox{\plotwidth}{!}{\includegraphics{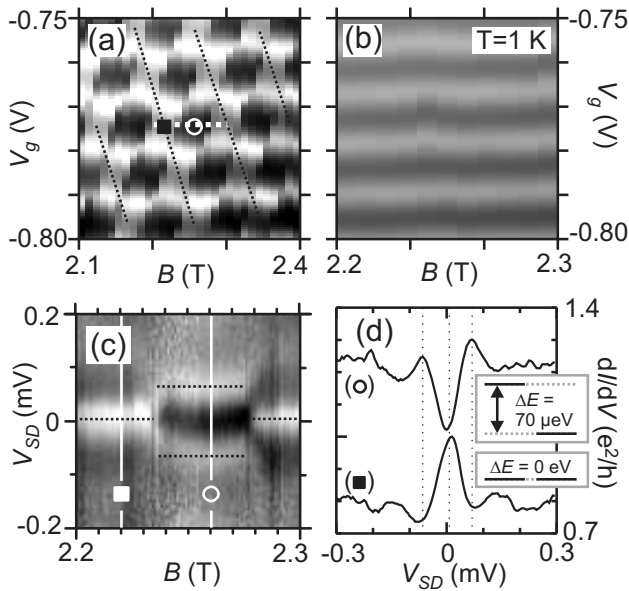}}
  \end{center}
  \caption{%
     Analysis of region (III) from Fig.~\ref{figStreifen}, $B \approx 2.2$~T :
     (a) The stripe pattern is formed by tiles of increased
         conductance between Coulomb blockade peaks over a wide parameter
         range.  The black dotted lines illustrate the stripes.
     (b) At $T=1$~K, the tile pattern in (a) has vanished.
     (c) Nonlinear magneto-conductance along the white dotted line in (a)
         in the Coulomb blockade valley at a fixed $V_g = -0.775$~V.
         Conductance maxima are highlighted with horizontal dotted
         lines.
     (d) Cuts as marked in (c).  Peak positions are highlighted
         with dotted lines at $V_{SD} = 0$~V and $\pm 70$~$\mu$V.
         The curves are offset for clarity.
    }
  \label{figKacheln}
\end{figure}

Figure \ref{figStreifen} shows an overview of the linear
conductance $G$ versus plunger gate voltage $V_g$ and
perpendicular magnetic field $B$.  For magnetic fields $B \gtrsim
1$~T a diagonal stripe pattern is clearly visible.  Similar
patterns were observed for several samples.  The striking
regularity of this pattern vanishes only for low fields and strong
coupling (upper left region in the Figure) where $G(B, V_g)$
becomes rather complicated.  This can be attributed to the
increased influence of disorder in this regime. Chaotic effects
typical for open quantum dots might also play a role.  We will
discuss the origin of the regular pattern - a modulated Kondo
effect - in Figures \ref{figWaben} and \ref{figKacheln} below but
will first focus on its periodicity.


\begin{figure} 
  \begin{center}
    \resizebox{\plotwidth}{!}{\includegraphics{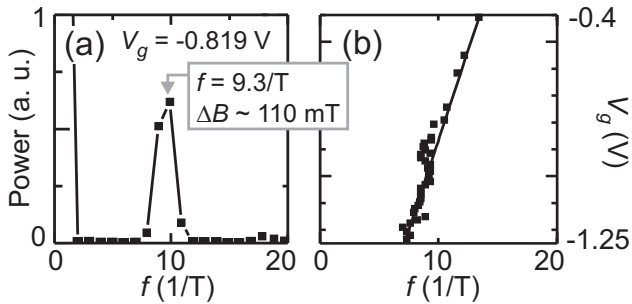}}
  \end{center}
  \caption{%
     Analysis of the $V_g$ dependent periodicity of the stripe
     pattern in Fig.~\ref{figStreifen} between $1.5$~T and $2.5$~T,
     where the pattern is most clearly visible:
     (a) Fourier transform of one line in Fig.~\ref{figStreifen}
     ($V_g = -0.819$~V) along the $B$ axis, showing a distinct peak
     corresponding to a periodicity of $\Delta B \approx 110$~mT.
     (b) Evaluations like in (a) show a roughly linear variation
     of the peak position with gate voltage $V_g$.
    }
  \label{figPeriodicity}
\end{figure}

The measured conductivity was Fourier transformed along the $B$
axis for several gate voltages $V_g$. Fig.~\ref{figPeriodicity}(a)
shows a typical result of the power spectrum obtained.  A clear
peak is observed at a frequency of $f=9.3$~T$^{-1}$ corresponding
to a periodicity of $\Delta B\approx 110$~mT.  Each such
transformation exhibits such a peak, from which we find a
periodicity varying from $\Delta B_1 = 130$~mT at $V_g = -1.2$~V
to $\Delta B_2 = 75$~mT at $V_g = -0.4$~V.  We have identified the
periodicity with the addition of one flux quantum to the dot:
$N_\phi = 1$ flux quanta $\phi_0$ added per stripe period lead to
dot diameters $d = 2 \sqrt{N_\phi \phi_0/\pi \Delta B}$ ranging
from 200~nm ($V_g=-1.2$~V) to 265~nm ($V_g=-0.4$~V) which is in
good agreement with the value of $d_{el} \approx 250$~nm estimated
above from the gate geometry. The smaller diameter for a more
negative plunger gate voltage is expected due to an increased
depletion.
From the calculated diameters we expect a change in the number of
electrons on the dot of $\Delta N = 87$ electrons which closely
resembles $\Delta N = 75$ electrons as known from Coulomb
blockade. The small difference could be easily attributed to the
uncertainty in the charge density $n$ and its unknown exact form
in the presence of charged top gates.

The addition of a flux quantum to a many electron system will
change its orbital and spin wave functions for magnetic fields
smaller than the extreme quantum limit \cite{McEuen-92,Ciorga-00}.
As an example, McEuen {\sl et al.}~\cite{McEuen-92} observed a
redistribution of electrons between different Landau levels within
their dot when a flux quantum was added.  Thus we can link our
stripe period to such redistributions of electrons.  In traces of
Coulomb peak positions $V_g(B)$ and amplitudes $G(B)$ we observe
consistent behavior.



To clarify the origin of the stripe pattern, we will now focus on
more detailed measurements in the fairly regular $B \sim 1$~T
regime as marked in Fig.~\ref{figStreifen} by (I).  The stripe
pattern is made up of regions of enhanced conductance in the
Coulomb {\sl blockade} regions (Fig.~\ref{figWaben}(a)), which
together with the Coulomb peaks form tiles of increased
conductance.  We have performed temperature and source-drain
voltage dependent measurements at the marked gate voltage and
magnetic field values to analyze this effect
(Fig.~\ref{figWaben}(c)). In the high conductance regions, a zero
bias peak is observed (circle in the figure).  It vanishes with
increasing temperature and disappears at a temperature of $T
\approx 1$~K. This zero bias peak is a clear signature of an
interaction effect. It can be attributed to the Kondo effect which
is illustrated by the characteristic temperature dependence in
Fig.~\ref{figWaben}(d). Due to the high and increasing background
conductance it is difficult to determine an exact Kondo
temperature $T_K$. After subtracting the exponential background
(triangle), the temperature dependence can be fit by the empirical
formula $G(T)=G_0(T_K'^2/(T^2+T_K'^2))^s$ with $T_K' =
T_K/\sqrt{2^{1/s}-1}$ \cite{Goldhaber-98b}, from which we extract
$T_K \approx 0.4$~K and $s \approx 1$. Although the value of $s$
strongly depends on the other fit parameters, it clearly deviates
from $s = 0.2$ characteristic for a spin $1/2$ system in the Kondo
regime. For the low conductance regions, the central Kondo peak is
suppressed (Fig.~\ref{figWaben}(c)).  Here the two small side
peaks at $\pm 170$~$\mu$eV which are also visible in the high
conductance trace become more prominent.  We expect them to be
related to a Kondo effect involving inelastic cotunneling through
excited states as discussed in \cite{Franceschi-01a}. This would
be roughly consistent with the level spacing $\Delta E \approx{}
150$~$\mu$eV stated above and also explains the background
conductance increasing exponentially with temperature in
Fig.~\ref{figWaben}(d) (triangle). In linear conductance
($V_{SD}=0$~V), the central Kondo peak appears as a high
conductance tile and the absence of a Kondo peak as a low
conductance tile.  Stripe and tile patterns can thus be explained
with a magnetically modulated Kondo effect.

In the $B \sim 1$~T regime, in some places the situation is not as
clear cut and deviations from the regular magneto-conductance
pattern are found. In Fig.~\ref{figWaben}(b) corresponding to
region (II) in Fig.~\ref{figStreifen}, we observe a more
honeycomb-like structure made up of narrow high-conductance lines
between two adjacent Coulomb blockade peaks instead of a
high-conductance tile.


Compared to the $B \sim 1$~T regime, for a higher magnetic field,
e.~g.~at $B \sim 2$~T, the stripe pattern is much clearer and
extremely regular (Fig.~\ref{figKacheln}(a), corresponding to
region (III) from Fig.~\ref{figStreifen}).  Different from the
former, the latter regime extends over a wide range of gate
voltage and magnetic field.  It consists of alternating tiles of
enhanced and suppressed conductance within the Coulomb blockade
regions like in region (I) discussed above.  The Kondo effect
observed here and thus the pattern itself vanishes totally with
increasing temperature (Fig. \ref{figKacheln}(b)), i.~e.~at
temperatures above $T \approx 0.5$~K only regular Coulomb blockade
resonances are observed. From measurements similar to the ones
presented in Fig.~\ref{figWaben}(d) we extract a Kondo temperature
of roughly $T_K \approx 0.2$~K.  To investigate the nature of the
Kondo physics found here, we examine the involved energy scales in
$V_{SD}$ dependent conductance measurements along the horizontal
dotted line in Fig.~\ref{figKacheln}(a) at a fixed gate voltage
(Fig.~\ref{figKacheln}(c)). We observe a central Kondo peak in
high conductance tiles which is abruptly split into two nearly
symmetric peaks in the low conductance region.  At the transition
the ground state of the dot and thus the nature of the Kondo state
changes. For the two different situations Fig.~\ref{figKacheln}(d)
shows two typical differential conductance measurements versus
$V_{SD}$ along the vertical lines marked in
Fig.~\ref{figKacheln}(c). In the high conductance/single peak
situation the ground state must be degenerate, i.~e.~the level
splitting must be $\Delta E=0$ {\em including} Zeeman energy.  In
the low conductance/split peak regime however, two states with
$\Delta E = 70$~$\mu$eV (also {\em including} Zeeman energy) must
be involved. This is illustrated in the insets in
Fig.~\ref{figKacheln}(d). The single peak situation is clearly
inconsistent with the simple model of one electron with spin
$S=\pm 1/2$ on a Zeeman split level from which a splitting $\Delta
E_Z = g_{GaAs} \mu_B B = 55$~$\mu$eV ($|g_{GaAs}|=0.44$) would be
expected.  Due to the low ratio $k_B T_K/E_Z = 0.3 < 1$ we should
be able to resolve such a splitting.  The Kondo states we observe
in our strong coupling case must be more complicated than a simple
hybridization between leads and one electron on the dot. For
example, a two-stage Kondo effect
\cite{Pustilnik-01,Hofstetter-02,Wiel-02} could be involved,
although we have no signature of a suppression of the Kondo effect
in the investigated temperature range.

Our result is different from the chessboard-like
magneto-conductance pattern investigated previously by Keller {\sl
et al.}~\cite{Keller-01} who observed an alternation between
Zeeman split spin-$1/2$ Kondo peaks in their high conductance
regions and no Kondo effect at all. We attribute this discrepancy
to a steeper confinement potential in their reactive ion etched
dot and to a lower Kondo temperature and coupling in comparison to
our system.





In conclusion, we explored Kondo physics in a large quantum dot
with strong coupling to the leads.  We observed a flux quantum
modulated Kondo effect over a wide parameter range in magnetic
field and in the number of electrons in the dot.  This Kondo
effect needs an explanation which goes beyond the classical
spin-$1/2$ Anderson model.


We thank F.~Hohls and U.~Zeitler for helpful discussions and help
with the measurement setup.  We acknowledge discussions with
M.~Pustilnik and A.~Tagliacozzo and financial support by DIP, TMR,
BMBF.  A.~D.~W. acknowledges support from SFB491.




\end{document}